# Weak ferrimagnetism and multiple magnetization reversal in α-$Cr_3(PO_4)_2$

A.N. Vasiliev[1], O.S. Volkova[1], E. Hammer[2], R. Glaum[2], J.-M. Broto[3], M. Millot[3,4], G. Nénert[5], Y. T. Liu[6], J.-Y. Lin[6], R. Klingeler[7], M. Abdel-Hafiez[8], Y. Krupskaya[8], A.U.B. Wolter[8], B. Büchner[8]

[1]Low Temperature Physics Department, Moscow State University, Moscow 119991, Russia
[2]Institute for Inorganic Chemistry, Bonn University, D-53121 Bonn, Germany
[3]Université de Toulouse; UPS, INSA, 143 Avenue de Rangueil, F-31400 Toulouse, France and Laboratoire National des Champs Magnétiques Intenses (LNCMI) -- CNRS UPR 3228, 143 Avenue de Rangueil, F-31400 Toulouse, France
[4]Department of Earth and Planetary Science, University of California-Berkeley, Berkeley, California 94720, USA
[5]Laue-Langevin Institute, Grenoble 38042, France
[6]Institute of Physics, National Chiao-Tung University, Hsinchu 30076, Taiwan
[7]Kirchhoff Institute for Physics, Heidelberg University, Heidelberg D-69120, Germany
[8]Leibniz Institute for Solid State and Materials Research, Dresden D-01069, Germany

The chromium(II) orthophosphate α-$Cr_3(PO_4)_2$ is a weak ferrimagnet with the Curie temperature $T_C$ = 29 K confirmed by a λ-type peak in specific heat. Dominant antiferromagnetic interactions in this system are characterized by the Weiss temperature $\Theta$ = - 96 K, indicating an intermediate frustration ratio $|\Theta|/T_C$ ~ 3. In its magnetically ordered states α-$Cr_3(PO_4)_2$ exhibits a remarkable sequence of temperature-induced magnetization reversals sensitive to the protocol of measurements, i.e. either field-cooled or zero-field-cooled regimes. The reduction of the effective magnetic moment 4.5 $\mu_B/Cr^{2+}$, as compared to the spin-only moment 4.9 $\mu_B/Cr^{2+}$, cannot be ascribed to the occurence of the low-spin state in any crystallographic site of the Jahn-Teller active $3d^4$ $Cr^{2+}$ ions. X-ray absorption spectra at the K-edge indicate divalent chromium and unravel the high-spin state of these ions at the $L_{2,3}$-edges. Weak ferrimagnetism and multiple magnetization reversal phenomena seen in this compound could be ascribed to incomplete cancellation and distortion of partial spontaneous magnetization functions of $Cr^{2+}$ in its six crystallographically inequivalent positions.

**Introduction**

The chromium (II) orthophosphate, α-$Cr_3(PO_4)_2$, belongs to the vast family of anhydrous phosphates of divalent metals $M_3(PO_4)_2$ with M = Mg, Ca, Cr - Zn. The numerous crystal structures met in this multitude differ in the interconnection patterns of the metal-oxide polyhedra being related to those in the naturally occurring minerals farringtonite [1], graftonite [2], and sarcopside [3]. The basic motif in their structures is that of chains of edge-sharing octahedra similar to the olivine structure type but with every fourth octahedron missing. This leaves trimers of octahedra containing two inequivalent cation sites, the central M1 site with a rather regular symmetry and two distorted M2 sites of the terminal octahedra.

Except, probably, manganese(II) orthophosphate, $Mn_3(PO_4)_2$ [4], the magnetic properties of the transition metals orthophosphates are well documented. Iron(II) orthophosphate, $Fe_3(PO_4)_2$ (sarcopside), orders antiferromagnetically (or ferrimagnetically) at $T_N$ = 44 K, with the Fe2 sites in each chain having opposite spin directions along the [100] axis, leaving the central Fe1 ion frustrated with no net magnetic moment [5]. Cobalt(II) orthophosphate, $Co_3(PO_4)_2$, shows the onset of antiferromagnetism at $T_N$ = 30 K, its magnetic structure being commensurate with the chemical unit cell with the magnetic cell doubled along the $a$-axis [6]. Nickel(II) orthophosphate, $Ni_3(PO_4)_2$, exhibits a three-dimensional magnetic ordering at $T_N$ = 17 K. Assumingly, there are ferromagnetic interactions within the $Ni_3O_{14}$ trimers which are coupled antiferromagnetically between them, giving rise to a purely antiferromagnetic structure [7]. Finally, copper(II) orthophosphate, $Cu_3(PO_4)_2$, was found to be antiferromagnetic with the Néel temperature $T_N$ = 22.5 K. The magnetic propagation vector (0 0 1/2) is referred to the triacute reduced chemical unit cell and the magnetic structure is collinear with equal moments of about 0.68 $\mu_B$ on each of the two Cu1 and Cu2, crystallographically inequivalent $Cu^{2+}$ ions [8]. Besides cited data on magnetism in monometallic orthophosphates $M_3(PO_4)_2$, an extensive information is available on the magnetic properties of the mixed metal orthophosphates, e.g. $CuNi_2(PO_4)_2$ [9] or $SrFe_2(PO_4)_2$ [10]. All of them exhibit long-range antiferromagnetic ordering at low temperatures, but the complexity of crystal structures hampers, usually, the parameterization of the magnetic subsystem.

One of the least studied members of this family, chromium(II) orthophosphate, $Cr_3(PO_4)_2$, can be found in two distinct crystallographic modifications [11, 12]. The crystal structure of the high-temperature $P2_1/n$ monoclinic phase β-$Cr_3(PO_4)_2$, stable in the temperature range 1250 - 1350 C, is close to that of farringtonite [1]. It contains interconnected zigzag chains of corner-sharing - edge-sharing elongated octahedra in the Cr2 – Cr1 – Cr1 – Cr2 sequence [12]. This phase experiences long-range magnetic ordering at $T_C$ = 36 K which is preceded by short-range correlation maximum at T ~ 60 K. The paramagnetic Weiss temperature in β-$Cr_3(PO_4)_2$ is negative, Θ = –165 K, indicating strong predominance of antiferromagnetic interactions. At variance with "standard" antiferromagnetic behavior, the magnetization in the magnetically ordered state of β-$Cr_3(PO_4)_2$, i.e. at T < $T_C$, rises with lowering temperature. This may indicate either incomplete cancellation of primarily antiparallel sublattice magnetizations or their canting due to the effects of magnetocrystalline anisotropy or Dzyaloshinskii – Moriya interaction.

The orthorhombic phase α-$Cr_3(PO_4)_2$ ($P2_12_12_1$, Z = 8, a = 8.4849(10) Å, b = 10,3317(10) Å, c = 14.206(2) Å) is stable in the temperature range between 1100 and 1250 °C, at lower temperature decomposition into CrP, $Cr_2O_3$, and $Cr_2P_2O_7$ is observed [11]. The unit cell parameters were determined at room temperature from X-ray powder diffraction data of a quenched sample. The local environment of the $Cr^{2+}$ ions is rather diverse as compared to the structure of β-$Cr_3(PO_4)_2$. Following the numbering scheme in [11] Cr1 to Cr5 show fourfold, distorted (roof-shaped) square-planar coordination with 1.96 ≤ d(Cr-O) ≤ 2.15 Å. Cr6 shows five oxygen ligands at 1.97 ≤ d(Cr-O) ≤ 2.29 Å.

The arrangement of the structural units in α-$Cr_3(PO_4)_2$ can be rationalized in terms of close-packed tubes (parallel to the crystallographic b-axis) with $Cr^{2+}$ on their inner surface and $PO_4$ tetrahedra on the outer surface as well as in the tube centers Fig. 1a [11]. The spatial arrangement of $Cr^{2+}$ ions is that of a double helix reminiscent of DNA molecule, as shown in Fig. 1b. The atomic arrangement within one tube is related to the one in adjacent tubes by pseudo $3_1$-screw axes. The local oxygen coordination of $Cr^{2+}$ ions exhibits five distorted square-planar groups $Cr1O_4$ to $Cr5O_4$ and a slightly distorted square-pyramidal unit $Cr6O_5$ (Fig. 1b). One chain formed by vertex-sharing of alternating $PO_4$ and $CrO_4$ units (containing Cr1, Cr2 and Cr3) winds along the pseudo $3_1$ axis. By edge-sharing between $Cr2O_4$ and $Cr5O_4$ this one is linked to a second type of chain. The latter consists of "dimers" $Cr4O_4$-$Cr6O_5$, which are linked via $Cr5O_4$ groups and $PO_4$ tetrahedra by vertex-sharing. This second chain winds around the $2_1$ screw axes parallel to the b-axis at the centre of the "tubes" described above (Fig. 1b). An unusual structural feature is the vertex-sharing between the almost orthogonal square-planes $Cr4O_4$ and $Cr5O_4$. Thus, a rather short distance d(Cr4-Cr5) = 3.08 Å) is formed. An unusually low average magnetic moment $\mu_{eff}$ = 4.28 $\mu_B$ per $Cr^{2+}$ ions and negative Weiss temperature Θ = –55 K were reported for α-$Cr_3(PO_4)_2$ [11]. In this paper we present the first extensive experimental study to elucidate the unexpected behavior of this compound.

**Experimental**

The present study of α-$Cr_3(PO_4)_2$ includes the sample preparation, the K and $L_{2,3}$ edges X-ray absorption spectroscopy (XAS), measurements of specific heat in the range 2 - 100 K, magnetic susceptibility measurements in the range 2 - 300 K, and pulsed magnetic field measurements up to 50 T.

*Sample preparation*

The α-modification of chromium(II) orthophosphate $Cr_3(PO_4)_2$ has been synthesized according to Ref. [11] from mixtures of $CrPO_4$ and Cr metal in the ratio 2:1 in evacuated silica ampoules at 1200°C (4 days) and quenched to room temperature. By chemical vapor transport (transport agent $I_2$, 1200 → 1100 °C, quartz ampoule) deep blue-violet single crystals of α-$Cr_3(PO_4)_2$ with edge-lengths up to several tenths of a millimeter have been obtained.

To avoid irreproducibility of results in different measurements due to the effects of magnetocrystalline anisotropy the necessary amounts of small single crystals were crushed into powder in the agate mortar and pressed into pellets. The material used for all measurements was selected under a microscope from crystals deposited by chemical vapour transport. Its X-ray powder diffraction pattern showed no traces of impurities.

*X-ray absorption spectroscopy*

The K edge and $L_{2,3}$ edges of chromium in α-$Cr_3(PO_4)_2$ were recorded at at EXAFS and HSGM beam lines, respectively, of the National Synchrotron Radiation Research Center in Taiwan. The metal K edge corresponds to excitation of 1s electrons to valence bond states localized on the metal. The energy and the shape of the X-ray absorption near edge structure characterize the local symmetry and the oxidation state of the metal ions. To the first approximation, the correlation between the energy of the K-edge and the valence state of chromium is linear [13]. The XAS spectra in α-$Cr_3(PO_4)_2$ and several reference compounds in different oxidation states [14] are shown in Fig. 2. The XAS K-edge in various compounds was defined *a*s an inflection point in corresponding spectra. The removal of one electron from the valence shell of chromium results in ~ 3 eV shift of the K- edge to higher energy. The dependence of the K-edge energy on the oxidation state of chromium is shown in Fig. 3. As expected from its chemical formula, the observed value of the K-edge in α-$Cr_3(PO_4)_2$ is a clear signature of divalent chromium.

The $L_{2,3}$ edges in XAS spectra correspond to excitations of 2p electrons to the partially unfilled 3d shell. XAS spectra taken at the transition metal $L_{2,3}$ edges are highly sensitive to both the valence and the spin states [15]. An increase of the valence state of the metal ion by one causes a shift of XAS $L_{2,3}$ edges by ~ 1 eV toward higher energy. This shift is due to a final state effect in the X-ray absorption process. The energy difference between a $3d^n$ ($3d^4$ for $Cr^{2+}$) and a $3d^{n-1}$ ($3d^3$ for $Cr^{3+}$) configurations is

$$\Delta E = E(2p^6 3d^{n-1} \to 2p^5 3d^n) - E(2p^6 3d^n \to 2p^5 3d^{n+1}) \approx U_{pd} - U_{dd} \approx eV$$

where $U_{dd}$ (resp. $U_{pd}$) is the Coulomb repulsion energy between two 3d electrons (between a 3d electron and the 2p core hole) [16]. Both energies are sensitive to the arrangement of the electrons on the d shell, i.e. to the spin state of metal.

In Fig. 4, a shift of the $L_3$ edge in divalent α-$Cr_3(PO_4)_2$ system to lower energy by approximately 1 eV as compared to $Cr_2O_3$ containing trivalent chromium is evidenced. Moreover, the detailed analysis of the Cr $L_{2,3}$ edges in α-$Cr_3(PO_4)_2$ allows one to suggest that it is very similar to the behavior of $CrF_2$ where the high-spin S = 2 ground state for the $3d^4$ electron configuration was firmly established [17]. The fine structure of the $L_{2,3}$ spectral features deserves a complementary analysis. In particular, we do not propose any interpretation for the appearance of extra peaks of moderate amplitudes $CT_{1-3}$ between $L_3$ and $L_2$ edges which were attributed to charge transfer from the ligand valence orbitals to the Cr 3d orbitals in $CrF_2$.

*Specific heat*

The temperature dependence of specific heat $C_p$ in α-$Cr_3(PO_4)_2$ is shown in Fig. 5. The obvious λ-peak at $T_C$ = 29 K indicates a second-order phase transition from the paramagnetic to the magnetically ordered phase. Although fluctuations yield a seemingly enlargement of the specific heat jump at $T_C$ so that the experimentally observed anomaly clearly overestimates the mean field result, the observed value $\Delta C_p$ = 10.5 J/mol K is far smaller than expected in the mean field theory:

$$\Delta C_p = \frac{5nRS(S+1)}{(S+1)^2 + S^2} = 57.5 \frac{J}{molK}$$

where n = 3 is the number of magnetically active ions in the α-$Cr_3(PO_4)_2$ chemical formula, R = 8.314 J/mol K is the universal gas constant and S = 2 is the spin-only moment in presumably high-spin state of $Cr^{2+}$ ions. This fact indicates that a large amount of magnetic entropy is released above $T_C$ due to short-range magnetic correlations. Under a magnetic field of 9 T, the λ-peak somewhat broadens and slightly seems to shift to higher temperatures. The upward shift of this anomaly could be associated with the presence of ferromagnetic exchange interactions in the system stabilized by the external magnetic field. The influence of the magnetic field on the specific heat in α-$Cr_3(PO_4)_2$ is illustrated by the inset to Fig. 4, with a $C_p/T$ vs. $T^2$ plot of the experimental data. Such a presentation usually allows one to separate ~ $T^3$ terms, assigned to phonons and three-dimensional antiferromagnetic magnons, from any other terms of lower dimensionality. Here however in the case of α-$Cr_3(PO_4)_2$, this procedure is hampered by a large Schottky-type anomaly, hardly sensitive to the magnetic field. The indifference of this feature to significantly strong magnetic field signals its non-magnetic origin.

*Magnetic susceptibility*

The temperature dependence of the magnetic susceptibility χ of α-$Cr_3(PO_4)_2$ taken at B = 0.1 T is shown in Fig. 6. The smooth increase of χ seen under decreasing temperature is followed by the abrupt jump of the signal at $T_C$ = 29 K in agreement with earlier observations [11], typical of a compound whose magnetization in the magnetically ordered state contains a ferromagnetic component. Below $T_C$ however, the magnetic susceptibility evidences a remarkable sequence of temperature-induced magnetization reversals sensitive to the protocol of measurements, i.e. either field-cooled (FC) or zero-field-cooled (ZFC) regimes, as shown in the inset of Fig. 6. Note that in the FC regime the magnetic susceptibility even shows a "diamagnetic" response at lowest temperatures, which rapidly disappears as the external field is increased above 0.1 T.

The temperature dependence of the reciprocal magnetic susceptibility $\chi^{-1}(T)$, shown in Fig. 7, indicates the predominance of antiferromagnetic interactions in α-$Cr_3(PO_4)_2$. In the 200 – 300 K range the experimental data can be fitted by the Curie-Weiss law with inclusion of the temperature independent term $\chi_0$, i.e.

$$\chi = \chi_0 + \frac{C}{T - \Theta} = \chi_0 + n\frac{N_A g^2 S(S+1)\mu_B^2}{3k_B(T - \Theta)},$$

where C and Θ are the Curie and Weiss constants, $N_A$, $\mu_B$, and $k_B$ are the Avogadro, Bohr and Boltzmann constants, g is the g-factor. In this framework, the paramagnetic $\chi_0$ = 1.6 ×$10^{-4}$ emu/mol is that of summation of diamagnetic and van Vleck contributions, the Weiss temperature Θ = –96 K is large and negative, and the effective magnetic moment $\mu_{eff}$ = $[ng^2S(S+1)]\mu_B$ = 4.50±0.05 $\mu_B$ is significantly smaller than the spin-only value 4.9 $\mu_B$ per $Cr^{2+}$ ion. The reduced value of the effective magnetic moment $\mu_{eff}$ is hardly attributable to the presence of an unquenched orbital magnetic moment which might reduce the effective g-factor. While there are neither X-band (~ 9 GHz) nor Q-band (~35 GHz) electron spin resonance (ESR) studies of "ESR-silent" non-Kramers $3d^4$ $Cr^{2+}$ ions, the measurements at very high frequencies (~ 90 – 440 GHz) provided an value of g = 1.98 for the g-factor of $Cr^{2+}$ in frozen aqueous solutions [18].

Below 200 K, the reciprocal susceptibility $\chi^{-1}(T)$ in α-$Cr_3(PO_4)_2$ deviates from linearity and unravels ferromagnetic interactions in the system through the temperature dependence of the effective Curie constant C = (χ – $\chi_0$)×(T – Θ) shown in Fig. 7. Evidently, the short-range magnetic correlations develop in this compound far above the magnetic ordering temperature in correspondence with a significant reduction of the jump $\Delta C_p$ in specific heat at $T_C$.

*High-field magnetization*

The field dependences of magnetization taken at several temperatures both below and above the magnetic ordering temperature $T_C = 29$ K are shown in Fig. 8. Note, that the remanent magnetization at $T < T_C$ strongly depends on the protocol of measurements, i.e. why the M(B) curve taken at 27 K shows negative magnetization at low fields. The magnetization loop in α - $Cr_3(PO_4)_2$ taken at 2 K is shown in the inset to Fig. 8. At 2 K, the remanence is 0.024 $\mu_B$/f.u. and the coercivity is 0.2 T signaling presence of weak ferromagnetism.

The field dependence of magnetization M(B) in α – $Cr_3(PO_4)_2$ has been measured in pulsed magnetic field up to B = 50 T at T = 2 K at LNCMI-Toulouse using concentric inductive pick-up coils on a powdered sample ( Fig. 9). The M(B) curve shows two almost linear segments at $B < B_1 \sim 5$ T and $B > B_2 \sim 30$ T, along with rather pronounced variations of M vs. B rate at $B_1 < B < B_2$. While the overall behavior of the magnetization is reminiscent of that at spin-flop and spin-flip transitions in antiferromagnets, the value of M ~ 6 $\mu_B$ at B = 50 T indicates that the system is far from the saturation. The extrapolation of the linear segment of the M(B) curve to the saturation magnetization value

$$M_{sat} = gS\mu_B = 11.9\mu_B$$

for g = 1.98 provides an estimate of the saturation field $B_{sat} \sim 130$ T. This is in rough correspondence with the estimation of the antiferromagnetic exchange interaction parameters which can be deduced from the negative value of the Weiss temperature $\Theta = -96$ K. From mean field theory $B_{sf} \sim (2B_aB_{sat})^{1/2}$ we can hence obtain an estimation of the effective field $B_a$ of magnetocrystalline anisotropy $B_a \sim 0.1$ T. We note, however, that the high-field magnetization data do not resemble a typical spin-flop phase behavior but indicate a more complex ferromagnetic-like phase since it does extrapolate to the finite magnetization M(B=0) ~ 3 $\mu_B$.

**Discussion**

Up to now, it is generally accepted that the presence of the Jahn-Teller ion $Cr^{2+}$ in an oxide environment is restricted to the case of the trirutile chromium(II) tantalate, $CrTa_2O_6$, which is an antiferromagnet with complex magnetic structure, the Néel temperature $T_N = 10.3$ K [19], and an effective magnetic moment $\mu_{eff}$ = 4.44 $\mu_B$ [20] (or 4.72 $\mu_B$ [19]). On the one hand, the discrepancy regarding the magnetic moment might already be taken as an indication on some uncertainty in the purity of the previously studied powder samples. On the other hand, a large series of very well defined, well crystallized and pure phosphates of divalent chromium are easily accessible by chemical vapour transport (e. g.: α-/β-$Cr_3(PO_4)_2$ [11, 12], $Cr_2P_2O_7$ [26]) or solid state reactions (e. g.: $SrCrP_2O_7$ [27], $BaCrSi_4O_{10}$ [28]). These compounds are well suited to study the electronic structure, chemical bonding and cooperative magnetic behavior related to the presence of $Cr^{2+}$ ions. Ligand-field spectra of the aforementioned compounds have already been reported [12, 28]. Since chemical bonding behavior of the polyatomic phosphate ion, $(PO_4)^{3-}$, is quite different from that of the oxide ion, one should expect for phosphates and silicates of $Cr^{2+}$ significant deviations from the physical and chemical properties reported for $CrTa_2O_6$ [19, 20]. The magnetic properties of α-$Cr_3(PO_4)_2$, i.e. weak ferrimagnetism and multiple magnetization reversals, are evidences for this expectation and of considerable interest themselves.

The phenomena of ferrimagnetism are associated with a partial cancellation of antiferromagnetically aligned magnetic sublattices with different values of magnetic moments and/or different temperature dependences of magnetization. It is frequently observed in compounds containing different magnetic ions and can be seen also in materials containing only one type of magnetic ions, which are in different valence states or crystallographic positions [21, 22]. In the latter case, the origin of weak ferrimagnetism lies in the difference of molecular fields acting on inequivalent magnetic sites [23].

In some ferrimagnets (in Néel's classification N-type ferrimagnets [24]) the total magnetization of a substance vanishes at a certain compensation temperature. Both above and below this temperature the magnetization of different sublattices prevails. In this case magnetization reversal can be observed. In a weak magnetic field (less than the field of coercitivity) the magnetization changes sign at the compensation temperature. Even the metastable "diamagnetic" state can be fixed by the magnetocrystalline anisotropy in a certain temperature range [25].

A similar, but even more complicated, situation is seen in α-$Cr_3(PO_4)_2$ containing six crystallographically independent positions for the magnetically active $Cr^{2+}$ ions. While the overall arrangement of magnetic moments in this compound is essentially antiferromagnetic, slight variations in temperature dependences of the six inequivalent spontaneous magnetization functions produce remarkable effects of weak ferrimagnetism and multiple magnetization reversals. The estimation of the effective field of magnetocrystalline anisotropy $B_a \sim 0.1$ T allows the observation of "diamagnetic" metastable states, as seen at lowest temperatures in the FC curve measured at $B = 0.1$ T. Unfortunately, the well developed procedure for the description of magnetization reversal phenomena in two-sublattices ferrimagnets [23, 25] can not be applied unequivocally to α-$Cr_3(PO_4)_2$ with six magnetic sublattices. Similarly, the significant reduction of the effective magnetic moment, i.e. 4.5 $\mu_B$ per $Cr^{2+}$ ion, in the paramagnetic state cannot be associated with the complex magnetization behavior in the magnetically ordered state. A possible explanation might be the assumption of direct overlap of the $d(z^2)$ orbitals of adjacent square-planar units $Cr4O_4$ and $Cr5O_5$. As already pointed out these two square-planar units are almost perpendicular to each other and linked by an oxygen atom. Via the rather short distance d(Cr-Cr) = 3.08 Å chemical bonding might become possible, as it is well documented for

$Cr_2SiO_4$ [29]. Thus, one should expect for Cr4 and Cr5 a reduced paramagnetic moment which corresponds to the three remaining unpaired electrons on these ions. Actually, a rough estimate of the average paramagnetic moment for a system containing four $Cr^{2+}$ ions with S = 2 and two with only S = 3/2 would lead to μ = 4.5 $μ_B$ per $Cr^{2+}$. At present, it is unclear whether the charge transfer effect seen in the XAS spectra of α-$Cr_3(PO_4)_2$ might also be related to the reduced paramagnetic moment. Further investigation by neutron diffraction experiments may help to gain a better understanding of these peculiar magnetic properties.

**Summary**

The chromium (II) orthophosphate α-$Cr_3(PO_4)_2$ is a rare case of a weak ferrimagnet based on a single transition metal in one oxidation state $Cr^{2+}$. The magnetic ordering at the Curie temperature $T_C$ = 29 K is confirmed by a λ-type peak in specific heat. Dominant antiferromagnetic interactions in this system are characterized by the Weiss temperature Θ = –96 K, indicating a frustration ratio $Θ/T_C$ ~ 3 reasonable for a three-dimensional magnetic entity. In the magnetically ordered state α-$Cr_3(PO_4)_2$ exhibits a remarkable sequence of temperature-induced magnetization reversals sensitive to the protocol of measurements, i.e. either field-cooled or zero-field-cooled regimes. The significant reduction of the effective magnetic moment 4.5 $μ_B/Cr^{2+}$, as compared to the spin-only moment 4.9 $μ_B/Cr^{2+}$, cannot be ascribed to the formation of the low-spin state in any crystallographic site of the Jahn-Teller active $3d^4$ $Cr^{2+}$ ions. . X-ray absorption spectra at the K-edge indicate divalent chromium and unravel the high-spin state of these ions at the $L_{2,3}$-edges. Weak ferrimagnetism and multiple magnetization reversal phenomena seen in this compound could be ascribed to incomplete cancellation and distortion of partial spontaneous magnetization functions of $Cr^{2+}$ in its six crystallographically inequivalent positions.

**Acknowledgements**

We acknowledge the support of the present work by Deutsche Forschungsgemeinschaft Grants DFG 486 RUS 113/982/0-1 and WO 1532/3-1, Russian Foundation for Basic Research Grants RFBR 09-02-91336, 10-02-00021, 11-02-00083. Part of this work was supported by EuroMagNET II at LNCMI-T facility and National Science Council of Taiwan (NSC98-2112-009-005-MY3).

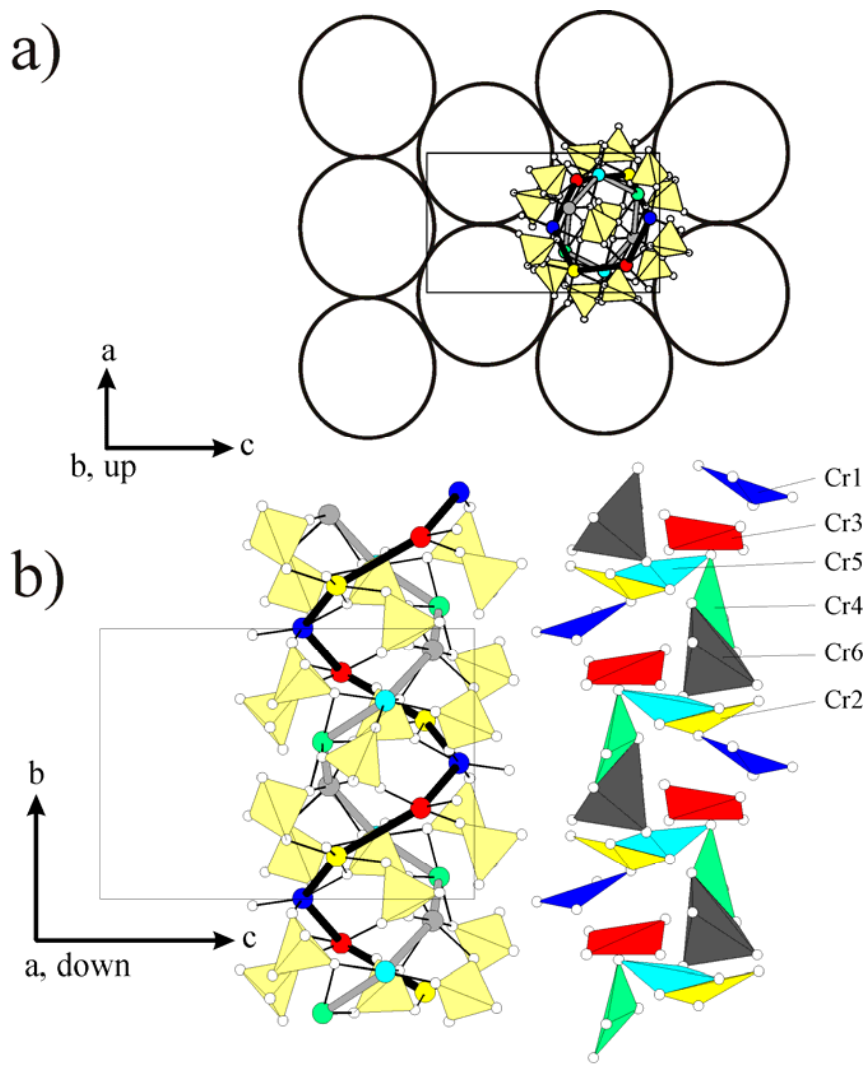

Fig. 1. Structural features of α-$Cr_3(PO_4)_2$. a) Close-packed arrangement of tubes from phosphate tetrahedra and $Cr^{2+}$ ions. b) Helical arrangement of the six independent $Cr^{2+}$ ions within a tube (right) and the same tube section with coordination polyhedra [$CrO_n$] (left). The phosphate groups $PO_4$ are represented by the light-yellow tetrahedra.

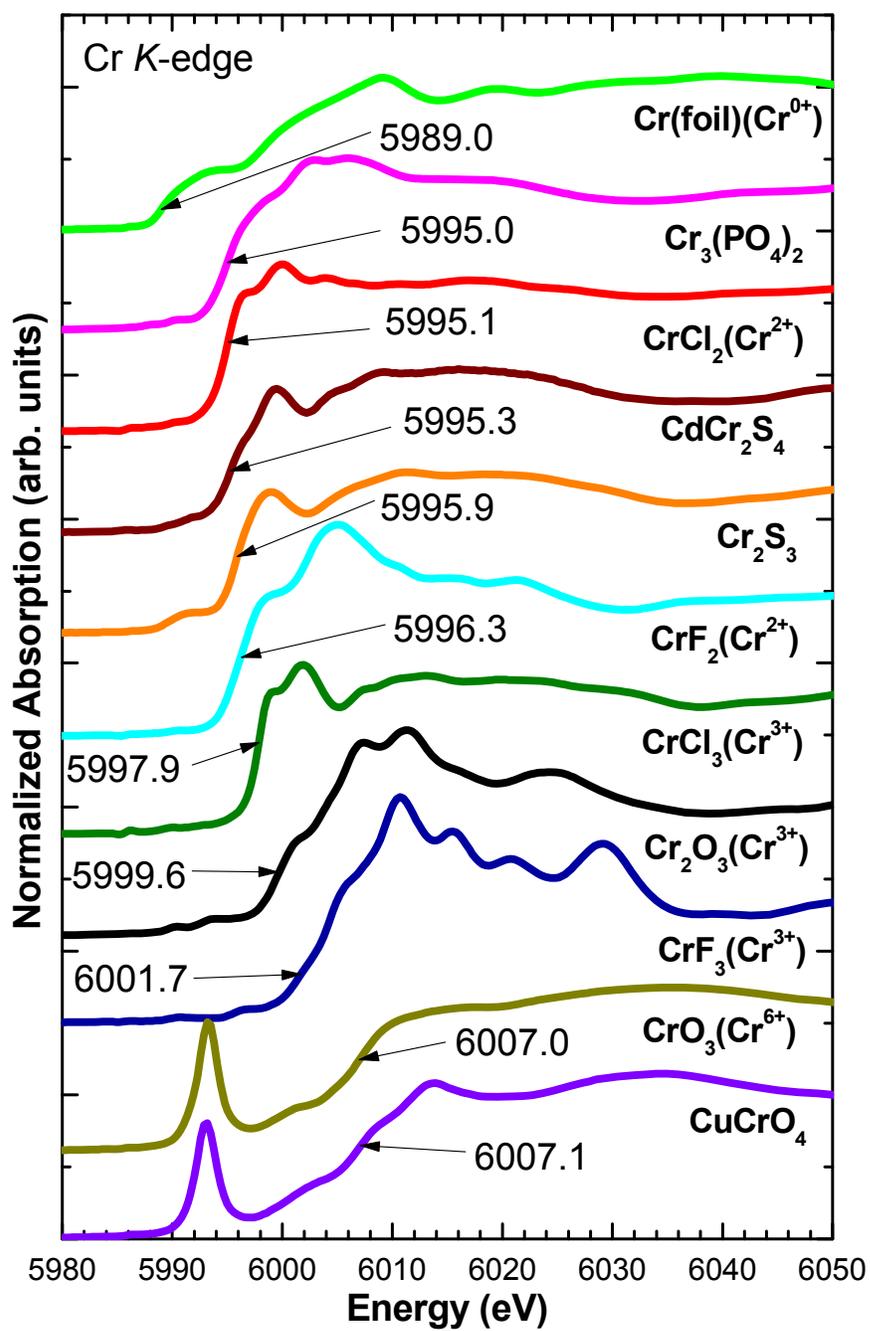

Fig. 2. The Cr K-edge X – ray absorption spectra in various chromium compounds taken at T = 300 K.

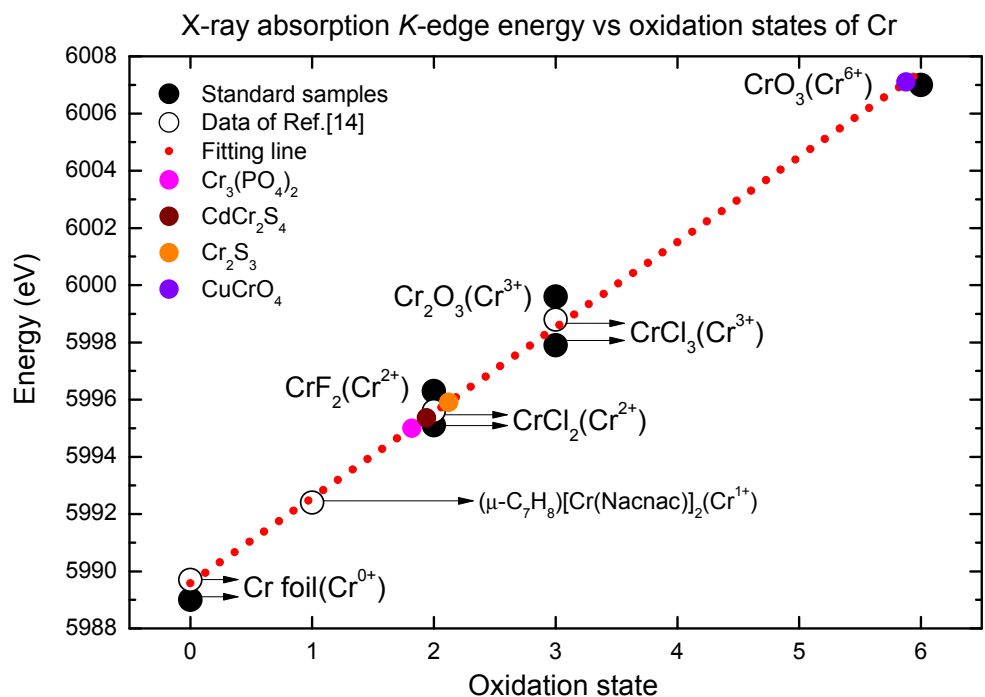

Fig. 3. The "linear" dependence of the K-edge energy on the oxidation state of chromium. The size of symbols corresponds roughly to the error bars.

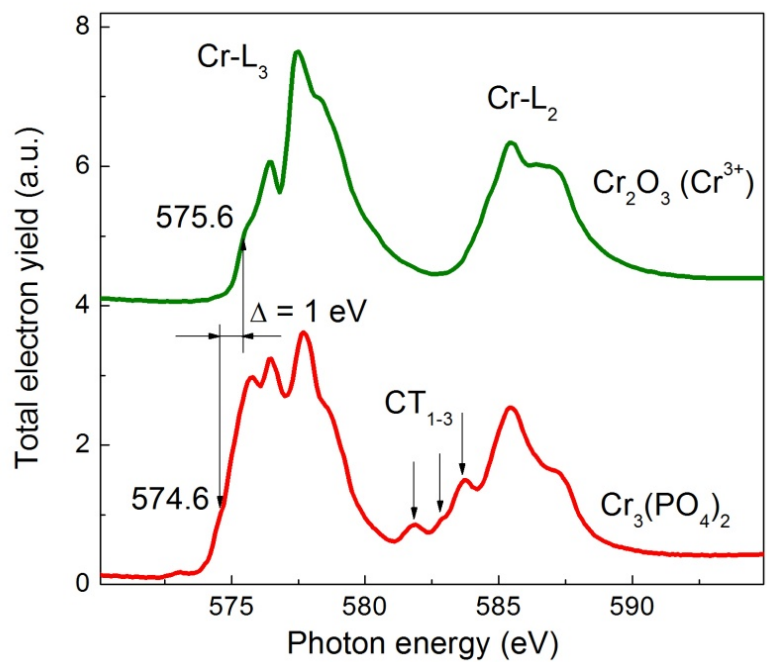

Fig. 4. The Cr $L_{2,3}$ edges X - ray absorption spectra in α-$Cr_3(PO_4)_2$ and $Cr_2O_3$ taken at T = 300 K. The $CT_{1-3}$ spectral features are ascribed tentatively to the charge transfer transitions.

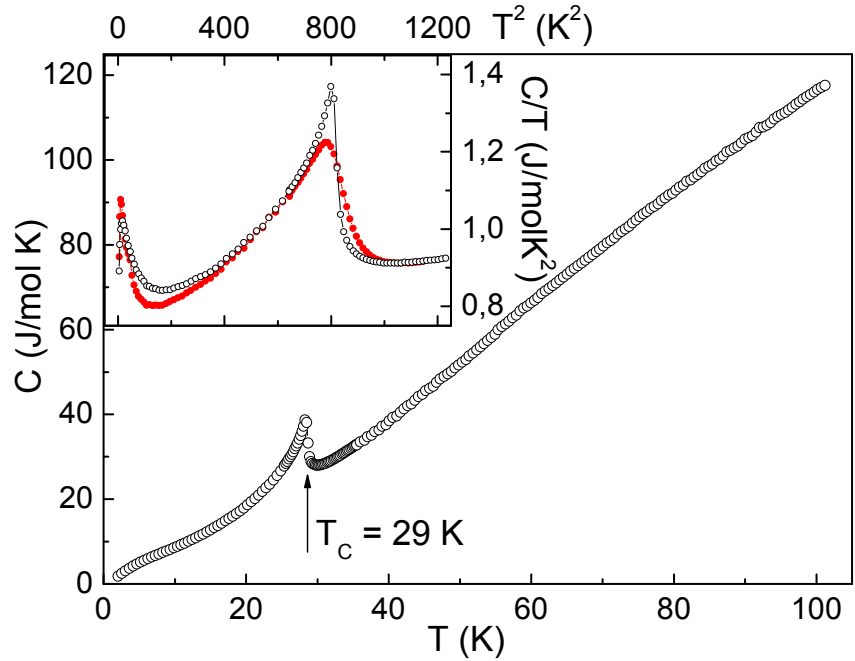

Fig. 5. The temperature dependence of the specific heat $C_p$ in α-$Cr_3(PO_4)_2$. Inset : Cp/T vs. $T^2$ curves taken at B = 0 and B = 9 T.

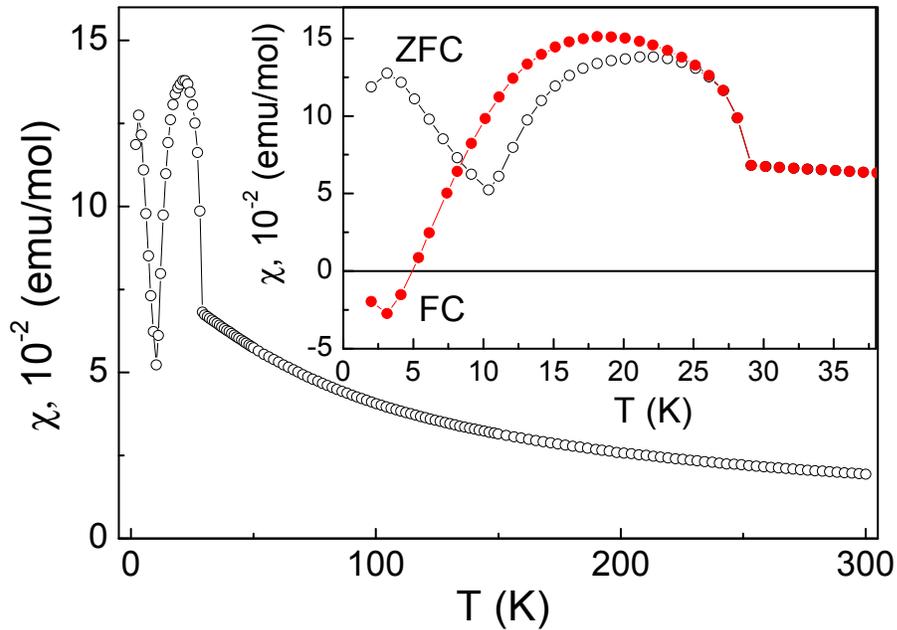

Fig. 6. The temperature dependence of magnetic susceptibility in α-$Cr_3(PO_4)_2$ taken at rising temperature at B = 0.1 T after cooling in zero field. Inset: enlarged portions of χ(T) curves taken with rising temperature after cooling at B = 0 (ZFC regime) and B = 0.1 T (FC regime).

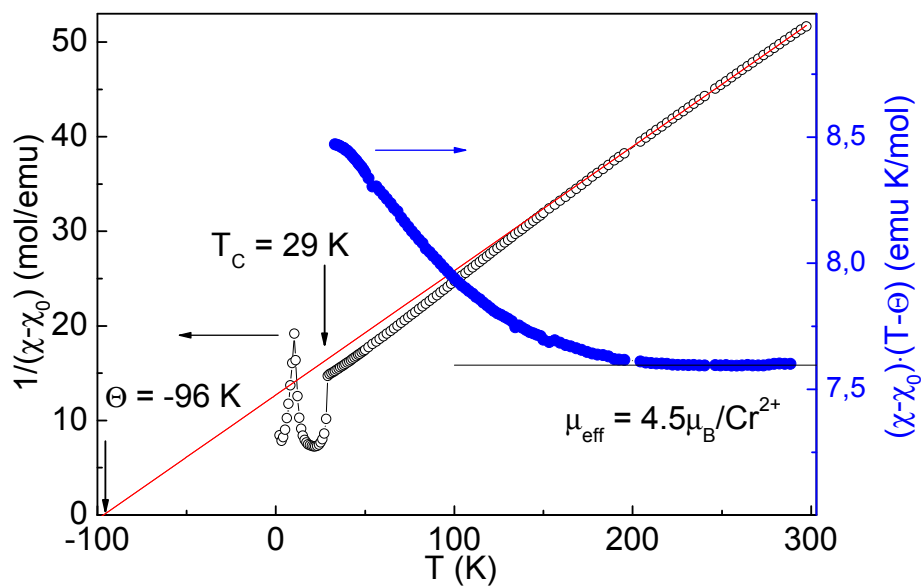

Fig. 7. The inverse magnetic susceptibility in α-Cr$_3$(PO$_4$)$_2$. The product $(\chi - \chi_0) \cdot (T - \Theta)$ indicates increasing relevance of ferromagnetic interactions when approaching the critical temperature $T_C$ = 29 K from above.

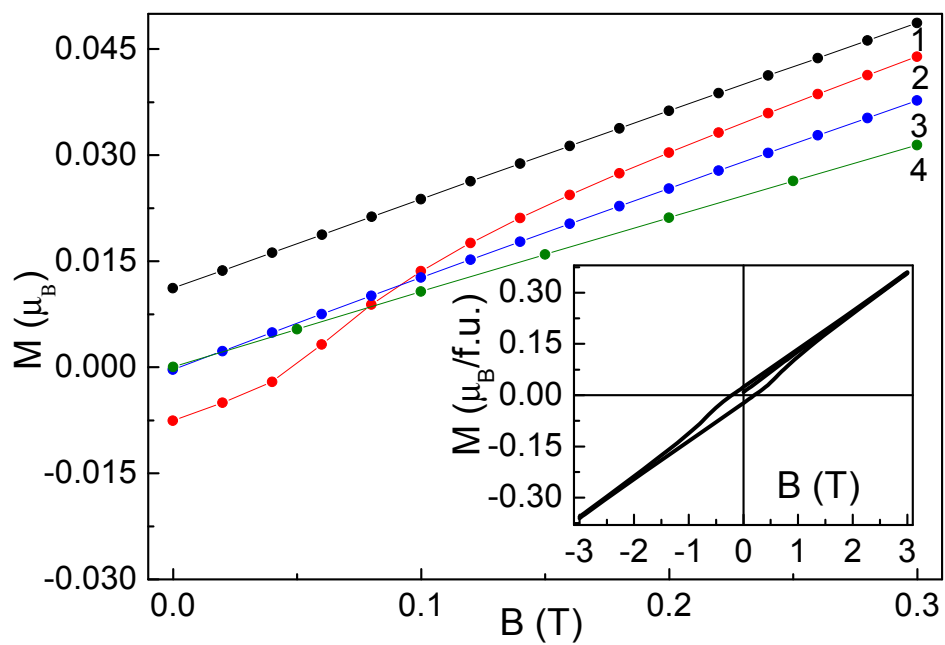

Fig. 8. The field dependences of magnetization in α-Cr$_3$(PO$_4$)$_2$ taken at selected temperatures both below and above the critical temperature $T_C$ = 29 K (1 – 25 K, 2 – 27 K, 3 – 29 K, 4 – 50 K). The inset represents the magnetization loop taken at 2 K.

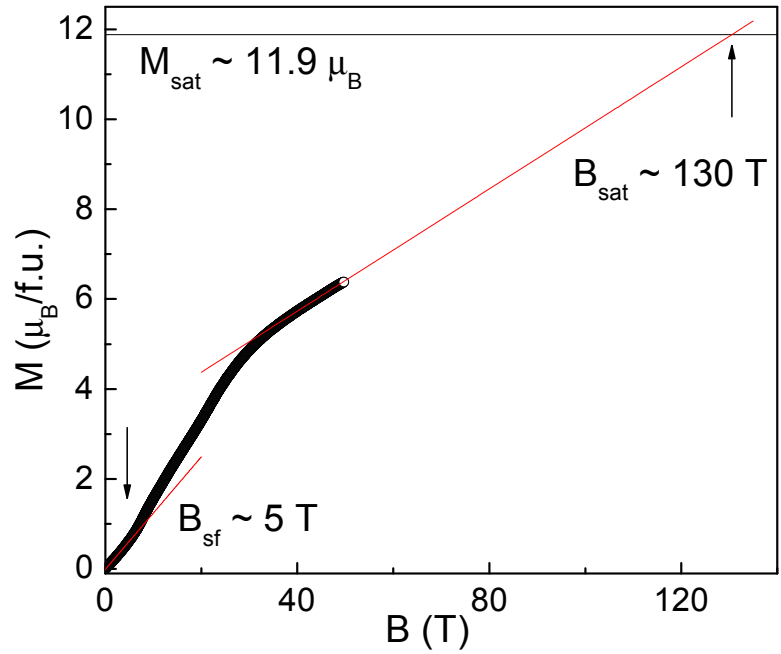

Fig. 9. The field dependence of magnetization in α-Cr$_3$(PO$_4$)$_2$ taken at T = 2 K. The solid lines are the extrapolations of the linear segments of the M(B) curve at B < B$_1$ and B > B$_2$. The horizontal line denotes value of saturation magnetization M$_{sat}$ for g – factor g = 1.98 μ$_B$ [19].